\documentclass[aps,prb,reprint,superscriptaddress]{revtex4-2}

\usepackage[colorlinks=true, linkcolor=blue, citecolor=blue,urlcolor=blue]{hyperref}
\newcommand{\scite}[1]{\textsuperscript{\cite{#1}}}

\usepackage{graphicx}
\usepackage{amsmath}
\usepackage{amssymb}
\usepackage{color}
\usepackage{float}
\usepackage{booktabs}
\usepackage{multirow}
\usepackage{booktabs}    
\usepackage{multirow}    
\usepackage{array}       
\usepackage{graphicx}    
\usepackage{siunitx}     
\usepackage{adjustbox}   

\newcommand{\SIQSE}{\affiliation{1}{Shenzhen Institute for Quantum Science and Engineering, Southern University of Science and Technology, Shenzhen 518048, Guangdong, China}}
\newcommand{\DPHY}{\affiliation{2}{Department of Physics, Southern University of Science and Technology, Shenzhen 518048, Guangdong, China}}
\newcommand{\IQA}{\affiliation{3}{International Quantum Academy, Shenzhen 518048, Guangdong, China}}
\newcommand{\GDKL}{\affiliation{4}{Guangdong Provincial Key Laboratory of Quantum Science and Engineering, Southern University of Science and Technology, Shenzhen 518048, Guangdong, China}}
\newcommand{\HFNL}{\affiliation{5}{
Shenzhen Branch, Hefei National Laboratory, Shenzhen 518048, China}}

\begin{document}
\title{A Low-Noise and High-Stability DC Source for Superconducting Quantum Circuits}

\author{Daxiong Sun}
\thanks{These authors contributed equally to this work.}
\affiliation{\SIQSE}\affiliation{\IQA}\affiliation{\GDKL}

\author{Jiawei Zhang}
\thanks{These authors contributed equally to this work.}
\email{zhangjw2022@mail.sustech.edu.cn}
\affiliation{\SIQSE}\affiliation{\IQA}\affiliation{\GDKL}

\author{Peisheng Huang}
\affiliation{\SIQSE}\affiliation{\IQA}\affiliation{\GDKL}

\author{Yubin Zhang}
\affiliation{\IQA}

\author{Zechen Guo}
\affiliation{\SIQSE}\affiliation{\IQA}\affiliation{\GDKL}

\author{Tingjin Chen}
\affiliation{\SIQSE}\affiliation{\IQA}\affiliation{\GDKL}

\author{Rui Wang}
\affiliation{\IQA}\affiliation{\GDKL}\affiliation{\DPHY}

\author{Xuandong Sun}
\affiliation{\IQA}\affiliation{\GDKL}\affiliation{\DPHY}

\author{Jiajian Zhang}
\affiliation{\IQA}

\author{Wenhui Huang}
\affiliation{\SIQSE}\affiliation{\IQA}\affiliation{\GDKL}

\author{Jiawei Qiu}
\affiliation{\IQA}

\author{Ji Chu}
\affiliation{\IQA}

\author{Ziyu Tao}
\affiliation{\IQA}

\author{Weijie Guo}
\affiliation{\IQA}

\author{Xiayu Linpeng}
\affiliation{\IQA}

\author{Ji Jiang}
\affiliation{\SIQSE}\affiliation{\IQA}\affiliation{\GDKL}

\author{Jingjing Niu}
\affiliation{\IQA}\affiliation{\HFNL}

\author{Youpeng Zhong}
\email{zhongyp@sustech.edu.cn}
\affiliation{\SIQSE}\affiliation{\IQA}\affiliation{\GDKL}\affiliation{\HFNL}

\author{Dapeng Yu}
\affiliation{\SIQSE}\affiliation{\IQA}\affiliation{\GDKL}\affiliation{\HFNL}

\date{\today}

\begin{abstract}
    With the rapid scaling of superconducting quantum processors, electronic control systems relying on commercial off-the-shelf instruments face critical bottlenecks in signal density, power consumption, and crosstalk mitigation. Here we present a custom dual-channel direct current (DC) source module (QPower) dedicated for large-scale superconducting quantum processors. The module delivers a voltage range of $\pm$7~V with 200~mA maximum current per channel, while achieving the following key performance benchmarks: noise spectral density of 20~nV/$\sqrt{\mathrm{Hz}}$ at 10~kHz, output ripple $<$500~$\mu$V$_{\mathrm{pp}}$ within 20~MHz bandwidth, and long-term voltage drift $<$5~$\mu$V$_{\mathrm{pp}}$ over 12~hours. Integrated into the control electronics of a 66-qubit quantum processor, QPower enables qubit coherence times of $T_1 = 87.6~\mu\mathrm{s}$ and Ramsey $T_2 = 5.1~\mu\mathrm{s}$, with qubit resonance frequency drift constrained to $\pm$40~kHz during 12-hour operation. This modular design is compact in size and efficient in energy consumption, providing a scalable DC source solution for intermediate-scale quantum processors with stringent noise and stability requirements, with potential extensions to other quantum hardware platforms and precision measurement.

\vspace{0.5em} 
\par
\vspace{0.5em} 
\end{abstract}
\maketitle

\section{Introduction}

\begin{figure*}[ht]
    \centering
    \includegraphics[width=0.9\textwidth]{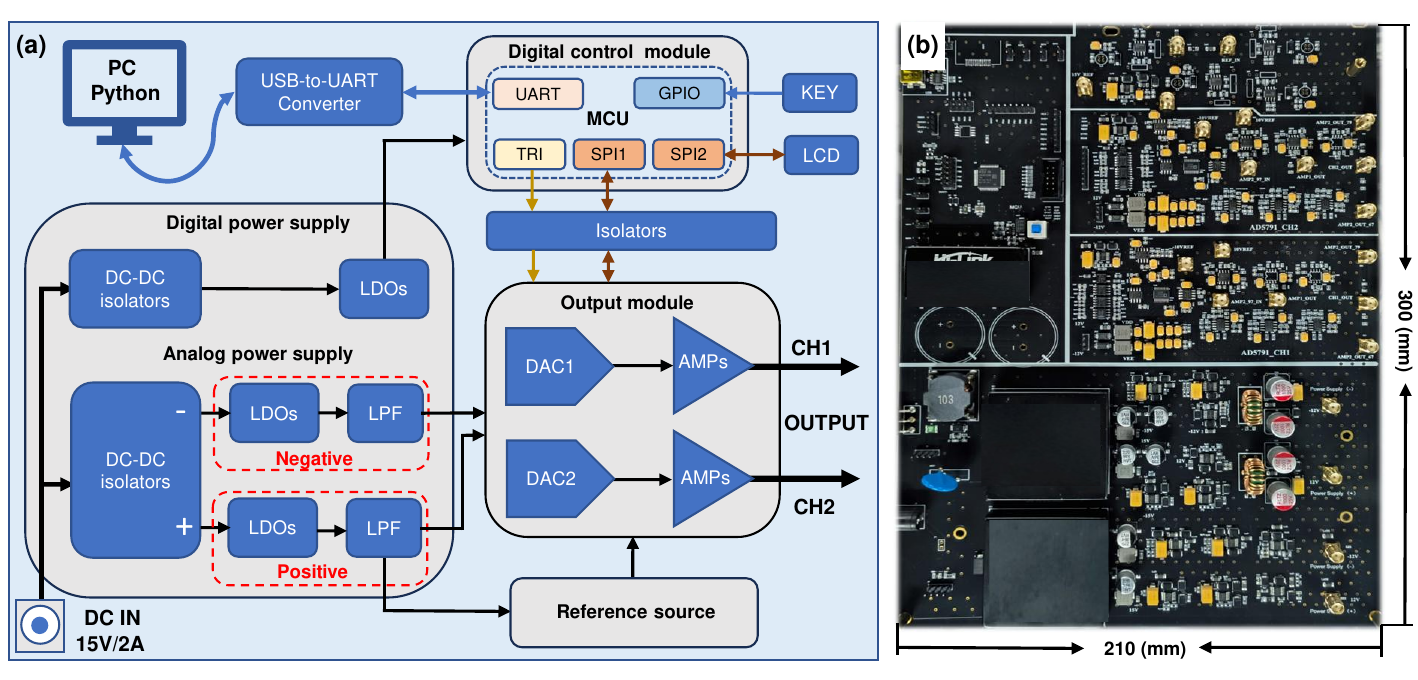}
    \caption{\textbf{\label{fig:1}System architecture of QPower.}
    (a) Block diagram of the hardware design. (b) Photograph of QPower.
    }
\end{figure*}

The rapid advancement of superconducting quantum circuits has culminated in processors integrating over one hundred qubits, marking significant progress toward practical quantum computing~\scite{IBM1000, xu2023digital, acharya2024quantum, gao2025establishing}. However, scaling these systems to thousands of qubits required for intermediate-scale quantum applications introduces formidable engineering challenges, not only for the quantum processors themselves, but also for the classical electronic control infrastructure~\scite{ball2021first,choi2023ibm,liang2018ultra,zhang2024m2cs}. Modern quantum processors rely on precision instrumentation for high-fidelity qubit manipulation and state readout, demanding stringent noise suppression and timing synchronization~\scite{li2019manipulation,kalfus2020high}. While commercial off-the-shelf electronic systems have enabled early-stage quantum experiments, their scalability limitations become apparent as qubit counts approach the thousands--a regime where conventional architectures face critical bottlenecks in signal density, power consumption, and crosstalk mitigation~\scite{rietsche2022quantum,ryan2017hardware,corcoles2019challenges}. This impending scalability barrier highlights an urgent need for integrated, application-specific control solutions that simultaneously optimize noise performance, power efficiency, and spatial footprint to meet the demands of next-generation quantum processors.

As superconducting qubits operate within the microwave frequency regime, microwave electronics are massively used to interface with their quantum dynamics~\scite{wang_hardware_2021-1,zhang2024m2cs,krantz2019quantum,yang2022fpga}. At the core of this infrastructure are arbitrary waveform generators (AWGs), which deliver nanosecond-level timing precision for coherent qubit control, and high-speed data acquisition (DAQ) modules that demodulate weak microwave readout signals to resolve quantum states~\scite{ding2024experimental,stefanazziQICKQuantumInstrumentation2022,xuQubiCExtensibleOpenSource2023,xuQubiCOpenSourceFPGABased2021,guoControlReadoutSoftware2019,linScalableCustomizableArbitrary2019,sunScalableSelfAdaptiveSynchronous2020}. Critically, frequency-tunable qubits and couplers themselves require microvolt-precision direct-current (DC) biasing to stabilize their operating points, a task as essential as waveform generation for maintaining quantum coherence~\scite{yang2022fpga, lisenfeld2023enhancing, grytsenko2024characterization, terai2003effects}. Furthermore, the microvolt-scale microwave responses from qubits demands a multi-stage amplification chain. Initial signal conditioning relies on cryogenic devices like Josephson parametric amplifiers (JPAs)~\scite{macklin2015near} or low-noise amplifiers (LNAs)~\scite{guo2025c}, followed by room-temperature amplification, each stage requiring stable DC biasing to maintain amplifier performance.  
Despite their foundational role, DC sources remain underappreciated compared to AWGs and DAQs. While commercial DC sources meet the noise and stability requirements for moderate-scale systems, scaling to thousands of qubits exposes critical bottlenecks: the cumulative cost, physical footprint, and power consumption of hundreds of discrete instruments become prohibitive~\scite{quinton2025quantum,ladd2010quantum,liang2018ultra}. Moreover, maintaining microvolt precision across thousands of channels exacerbates integration challenges, as crosstalk and ground loops degrade performance despite individual instruments meeting specifications.

\begin{table}[!h]
\caption{\textbf{Overview of key metrics of QPower.}} 
\label{table1}
\centering
\resizebox{0.9\linewidth}{!}{ 
\begin{tabular}{lc}
\toprule
\hline
Performance    & Value       \\
\midrule
Output      & 2      channels             \\
Output voltage range      & ±7        V    \\
Output current limit      & 200        mA    \\
Maximum power consumption   & 30       W      \\
Minimum step    & 15             $\mu$V                \\
Ripple    & \textless500        $\mu$V (20~MHz bandwidth)   \\
Long-term voltage drift    & \textless5        $\mu$Vpp (over 12 hours)    \\
Channel-to-channel crosstalk    &  $< 0.3$~ppm            \\
Low-frequency noise    & \textless20        nV/$\sqrt{\text{Hz}}$(@10 kHz)    \\
High-frequency spurious    & \textless$-95$       dBm (9~kHz~$\sim$~200~MHz)   \\
\hline
\bottomrule
\end{tabular}
}
\end{table}

To overcome these limitations, we introduce QPower -- a modular, high-density DC source architecture optimized for scalability. By co-integrating precision biasing for qubits and cryogenic amplifiers within a unified platform, QPower minimizes cost, size and power consumption (15~W per channel), while ensuring the low-noise and high-stability performance required for quantum processors. QPower adopts a distributed reference topology with LTZ1000ACH voltage references~\scite{yang2022fpga, analog_devices_LTZ1000}, achieving ripple noise below 500~$\mu$V$_{\mathrm{pp}}$. Hybrid regulation combining low-dropout regulators (LDOs) and active current sharing enhances output stability, while a phase-compensated amplifier array delivers 200~mA drive current with a 64.8$^\circ$ phase margin. Implemented in 2U chassis modules (2 channels/board), the design achieves a long-term drift below 5~$\mu$V$_{\mathrm{pp}}$, channel-to-channel crosstalk below 0.3~ppm, high-frequency spurious signals below $-95$~dBm (9~kHz~$\sim$~200~MHz), and low-frequency noise below 20~nV/$\sqrt{\text{Hz}}$ at 10~kHz. Key performance metrics are summarized in Table~1. Experimental validation on a 66-qubit processor~\scite{huang2025exact} demonstrates that QPower reliably supports intermediate-scale quantum systems, achieving $T_1 = 87.6 \pm 2~\mu\text{s}$ and $T_{\rm 2,ramsey} = 5.1 \pm 0.4~\mu\text{s}$ with $<$80~kHz peak-to-peak frequency drift over a 12-hour period.

\begin{figure*}[ht]
    \centering
    \includegraphics[width=0.9\textwidth]{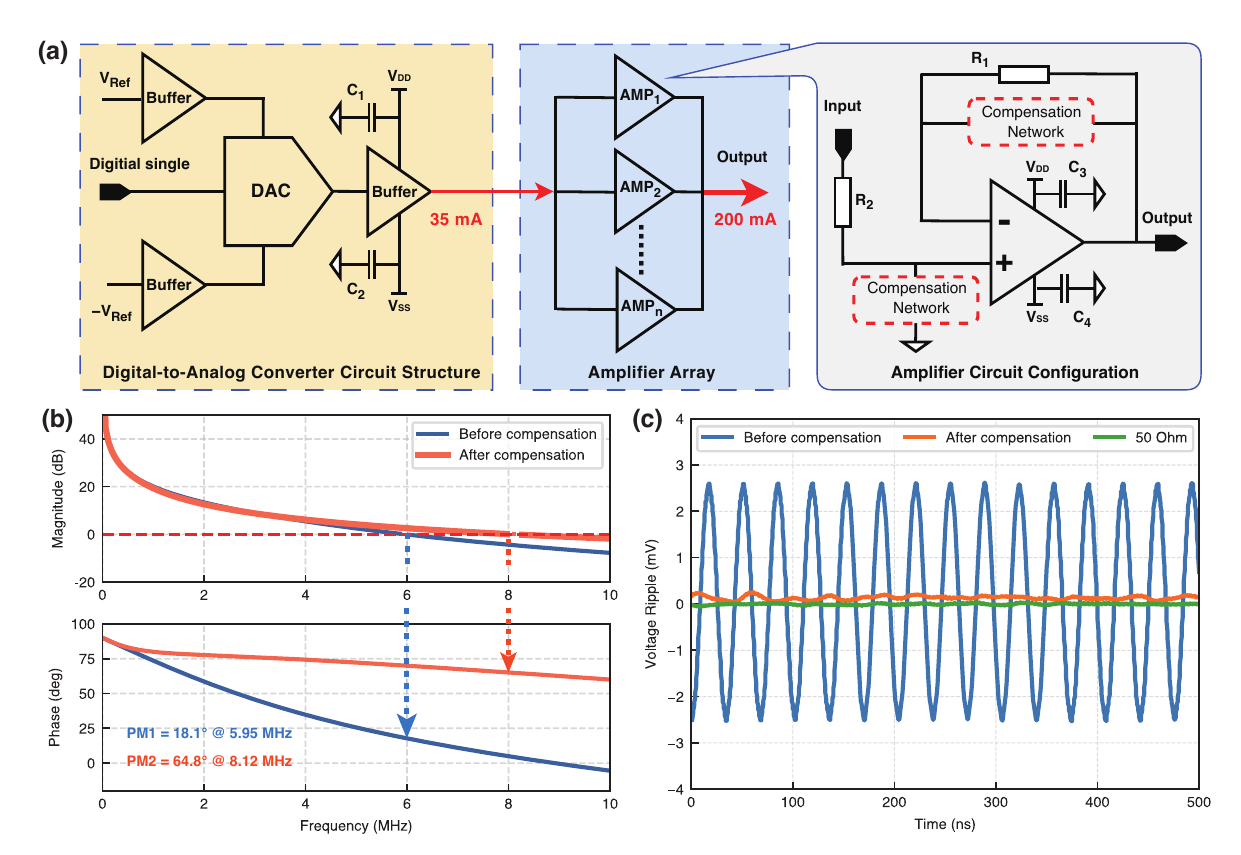}
    \caption{\textbf{QPower output design optimization.}
   (a) Schematic of the output design. The yellow region marks the DAC and buffer amplifier stage, the blue region shows the amplifier array, and the gray region highlights the compensation network. (b) Simulated open-loop gain and phase response of the output design. Phase margin (PM) increases from 18.1$^\circ$ (PM1) to 64.8$^\circ$ (PM2) after compensation, with the unity-gain bandwidth extended from 5.95~MHz to 8.12~MHz. (c) Measured output ripple before and after compensation. The blue trace shows ripple before compensation (5~mV$_\mathrm{pp}$), while the green trace shows post-compensation results (500~$\mu$V$_\mathrm{pp}$). The yellow trace indicates the oscilloscope noise floor under a 50~$\Omega$ load.
    }
     \label{fig:2}
\end{figure*}

\begin{figure*}[t]
    \centering
    \includegraphics[width=0.9\textwidth]{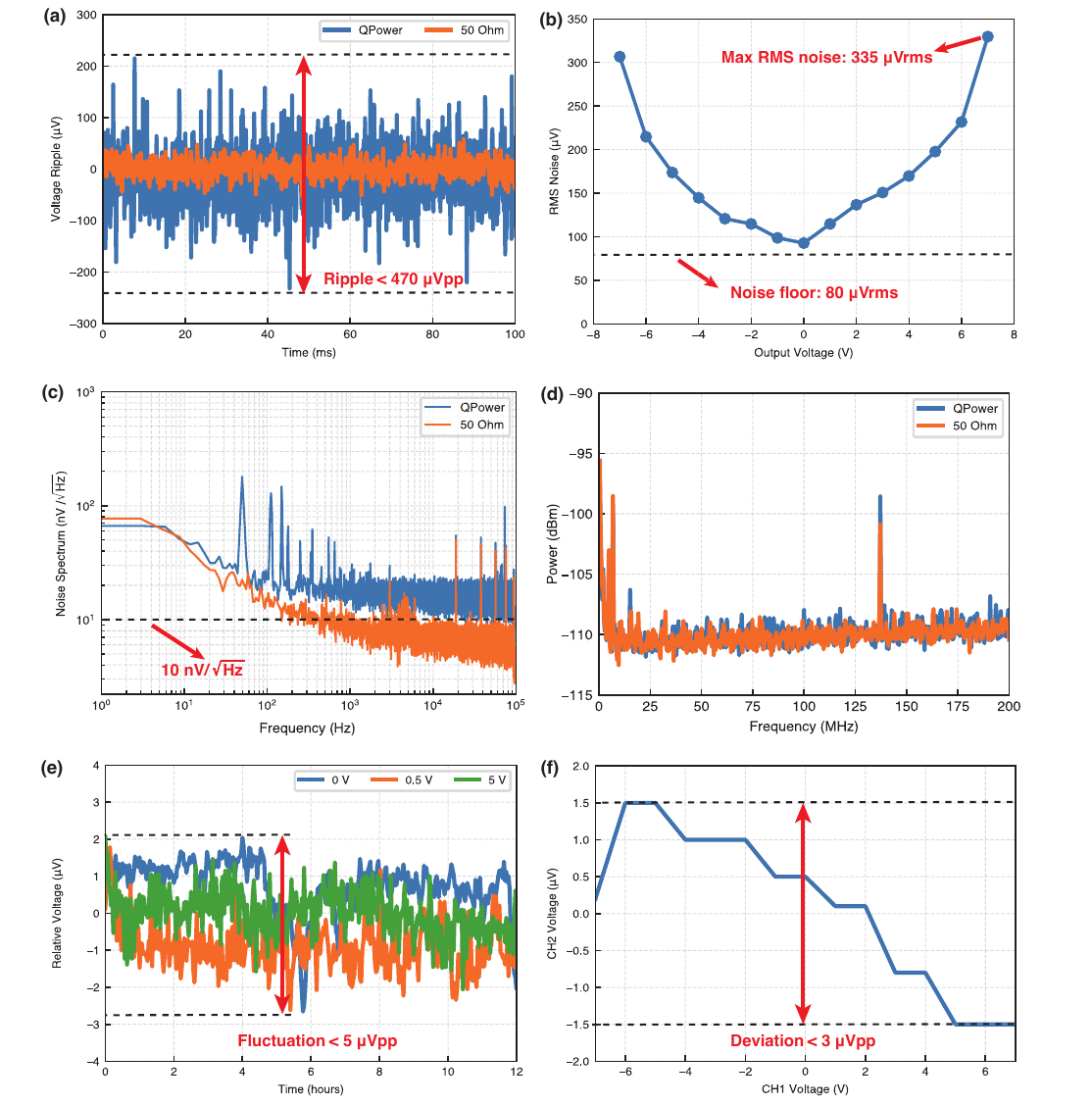}
        \caption{\textbf{Electronics performance.} (a) DC output ripple benchmark. Within the  DC~$\sim$~20~MHz bandwidth, the DC output ripple remains below 500~$\mu$V. The oscilloscope noise floor (approximately 100~$\mu$V) is verified using a 50~$\Omega$ load. (b) RMS noise benchmark. The root-mean-square (RMS) noise varies with output voltage, with a maximum value of 335~$\mu$V$_\mathrm{rms}$ at 7~V output. The instrument noise floor is approximately 80~$\mu$V$_\mathrm{rms}$. (c) Low-frequency noise spectrum of the DC output, showing a noise floor of 20~$\textrm{nV}/\sqrt{\textrm{Hz}}$ at 10~kHz. The instrument noise floor, approximately 10~$\textrm{nV}/\sqrt{\textrm{Hz}}$, is validated with a 50~$\Omega$ load. (d) High-frequency noise spectrum of the DC output. The noise remains below $-95$~dBm across the 9~kHz~$\sim$~200~MHz bandwidth. The instrument noise floor, approximately $-110$~dBm, as well as the peak at 137 MHz, is confirmed using a 50~$\Omega$ load. (e) Voltage fluctuation over 12 hours with the DC output set to $0$~V, $0.5$~V, and $5$~V, respectively. The peak-to-peak fluctuation is less than 5~$\mu$V. (f) Channel-to-channel crosstalk: The peak-to-peak voltage deviation at Channel 2 (CH2) remains below 3~$\mu$V$_{\text{pp}}$ when the Channel 1 (CH1) output is swept from $-7$~V to $+7$~V.}
    \label{fig:3}
\end{figure*}

\section{Electronics design and performance}

The hardware architecture of QPower is shown in Fig.~\ref{fig:1}(a), comprising four functional parts: a digital control unit based on a microcontroller (MCU), a power supply unit employing LDOs, an output stage with a high-resolution digital-to-analog converter (DAC) and amplifier arrays, and a precision reference voltage source. The physical prototype, measuring 210~mm in length and 300~mm in width, is presented in Fig.~\ref{fig:1}(b).

The digital control unit is responsible for coordinating system operation and regulating output voltages across all channels. To meet the demands of precise voltage control and flexible system access, both remote and local operation are supported. The control architecture integrates three communication interfaces -- Universal Asynchronous Receiver-Transmitter (UART), General Purpose Input/Output (GPIO), and Serial Peripheral Interface (SPI) -- managed via a tri-state bus controller. UART provides communication with a host computer for command input and status monitoring. GPIO allows manual adjustment of parameters on-site. The dual-channel SPI bus simultaneously controls the DAC and Liquid Crystal Display (LCD), enabling voltage updates and system feedback.
To ensure reliable data transmission in multi-device configurations, the SPI interface incorporates tri-state logic, which places inactive devices in a high-impedance state to avoid signal contention. During operation, the MCU executes embedded algorithms to process remote commands received via UART and local inputs from GPIO. These are converted into digital control signals and transmitted to the DAC via SPI for high-precision analog voltage generation. Qpower supports both remote operation via a host computer and local manual adjustment, ensuring adaptable control of the system across various operational scenarios.

The power supply part forms the energy backbone of the system, delivering stable power to both digital (MCU, peripherals) and analog (DAC, reference source, amplifier arrays) subsystems through a three-stage noise suppression architecture. First, front-end isolation achieves complete domain separation using dual DC-DC converters: one steps down the external +15~V input to 5~V for digital control circuits, while the other generates isolated ±15~V outputs for analog subsystems, complemented by optical isolators in control signal paths to maintain electrical isolation. Second, mid-stage regulation employs LT3045/LT3094 LDOs in serial–parallel topology, leveraging their high PSRR ($>90$~dB at 1~kHz) and low noise ($<0.8$~\textmu V\textsubscript{rms}) characteristics to condition the ±15~V rails for precision analog components. Finally, post-stage filtering implements a $\pi$-type Butterworth low-pass network (100~Hz cutoff, 20~dB attenuation at 1~kHz) to suppress residual high-frequency noise, completing the multi-level noise control strategy that ensures $<20$~nV/$\sqrt{\text{Hz}}$ output noise floor at 10~kHz.

The output part uses a cascade architecture of a 20-bit high-precision DAC (AD5791) and an instrumentation amplifier (AD8676), whose transfer function can be expressed as~\scite{wang2023design}:

\begin{equation} \label{eq:2-1}
V_{\rm OUT} = \frac{(V_{\rm REFP} - V_{\rm REFN}) \times D}{2^{20} - 1} + V_{\rm REFN},
\end{equation}
where $ V_{\text{REFP}} $ is the positive reference voltage, and $ V_{\text{REFN}} $ is the negative reference voltage, $ D $ is the 20-bit digital code written to the DAC register, with a range from 0 to $ 2^{20} - 1 $. The reference voltage range of the DAC, $ V_{\text{REFP}} - V_{\text{REFN}} $, determines the dynamic range of the output voltage. The output voltage $ V_{\text{OUT}} $ is obtained by linearly mapping the digital code $ D $ to the reference voltage range.

While the design fulfills the required voltage conversion functionality, practical implementation must address the limited driving capability of the AD5791 DAC, which exhibits a 3.4~k$\Omega$ output impedance causing significant signal attenuation in low-impedance or high-capacitance applications. To resolve this, we implement a two-stage architecture combining low-noise operational amplifier arrays and power amplification: First, the initial buffer stage (yellow box in Fig.~\ref{fig:2}(a)) \scite{egan201020} provides high-precision signal conditioning and noise suppression, though its maximum driving current of 35~mA proves inadequate for superconducting quantum system requirements. To enhance current capacity, we cascade a power amplifier array (blue area in Fig.~\ref{fig:2}(a)), boosting the maximum drive current to 200~mA -- achieving performance parity with commercial precision DC sources. This current enhancement introduces potential stability challenges, as the added amplification stage reduces phase margin (PM), risking self-oscillation \scite{de2015hysteresis, jenkins2013self}. Through careful phase margin optimization within the stable range of 45$^\circ$ to 70$^\circ$~\scite{pandiev2023stability,safonov1977gain,morroni2009design}, we implement a compensation network (gray box in Fig.~\ref{fig:2}(a)) that increases the unity gain bandwidth from 5.95~MHz to 8.12~MHz and improves PM from 18.1$^\circ$ to 64.8$^\circ$, as verified by LTspice simulations (Fig.~\ref{fig:2}(b)). Experimental validation (Fig.~\ref{fig:2}(c)) confirms the compensation eliminates self-oscillation, reducing output ripple from 5~mV\textsubscript{pp} to 500~\textmu V\textsubscript{pp}, aligning precisely with simulation predictions.

\begin{figure*}[t]
    \centering
    \includegraphics[width=0.9\textwidth]{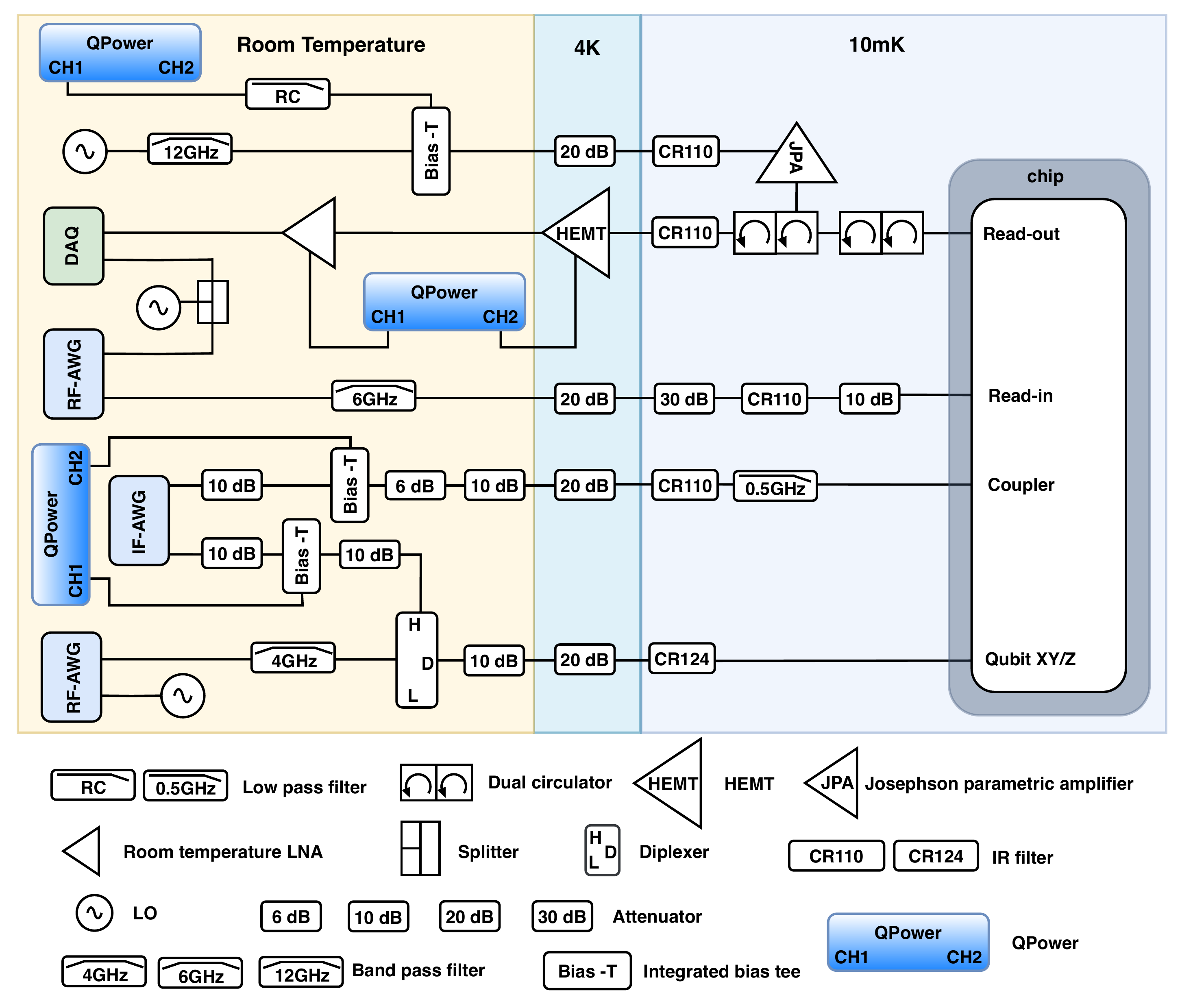}
    \caption{\textbf{Schematic of the control and readout setup for superconducting qubits.} 
    The M$^2$CS RF-AWG module generates XY-axis microwave driving signals, while the Z-axis control signals are synthesized through coordinated operation of the IF-AWG module and QPower, collectively enabling quantum bit manipulation. In the readout pathway, 6~$\sim$~8\ \text{GHz} microwave probe pulses generated by the RF-AWG module are injected into the quantum chip via cryogenic transmission lines. The returning quantum state response signals undergo isolation through circulators before being sequentially amplified by the JPA for primary parametric amplification, followed by secondary amplification through a HEMT, and tertiary gain enhancement via room-temperature low-noise amplifiers. The processed signals are ultimately demodulated and digitized by a M$^2$CS DAQ module. Three QPower modules are employed to provide the DC bias for the JPA, HEMT, and room-temperature LNA, a qubit, and a tunable coupler.}
    \label{fig:4}
\end{figure*}

\begin{figure*}[t]
    \centering
    \includegraphics[width=0.9\textwidth]{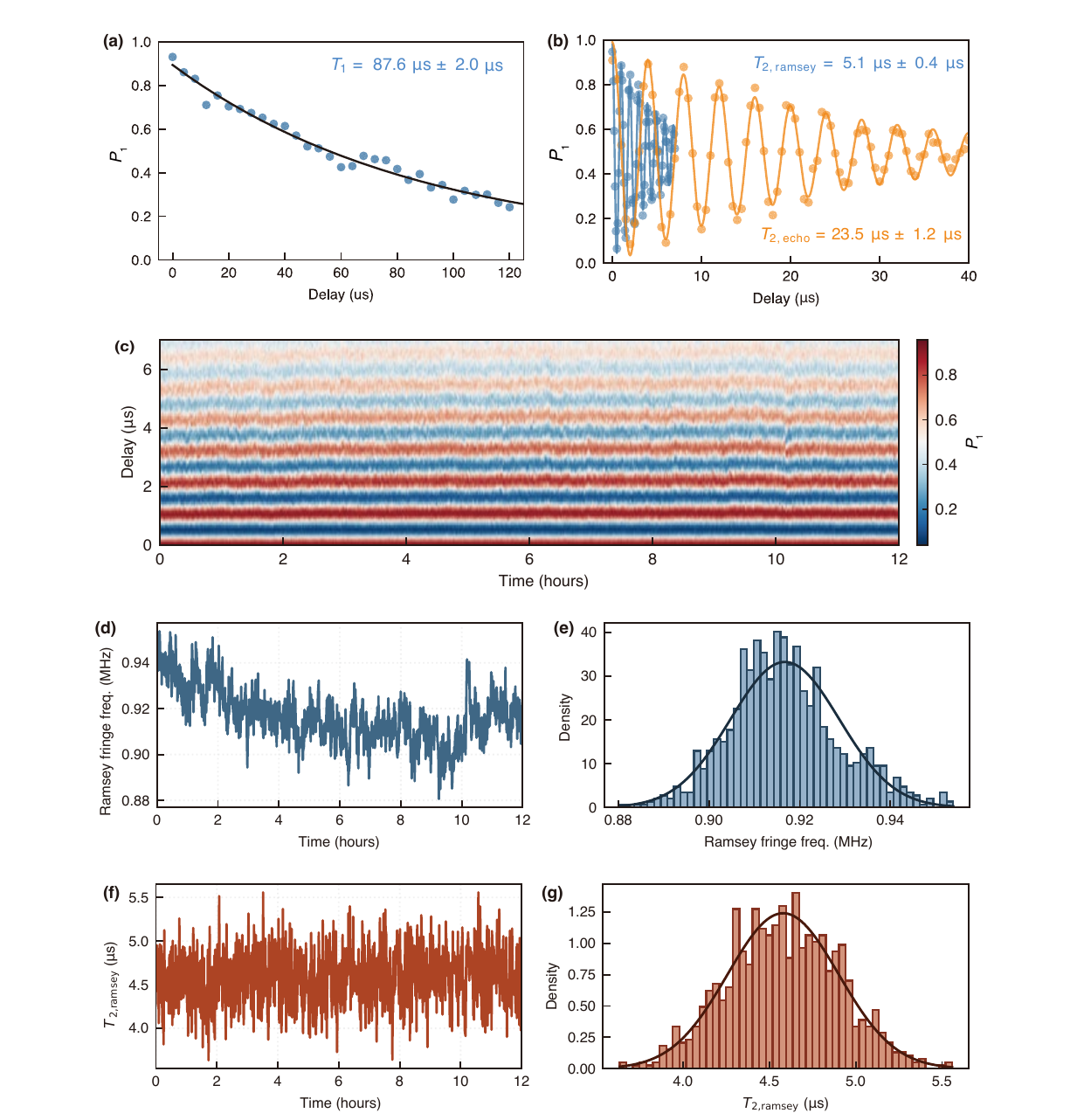}
    \caption{\label{fig:5}\textbf{Benchmarking QPower with superconducting qubits.}
    (a), (b) Qubit coherence benchmark showing $T_{1} = 87.6 \pm 2.0~\mu\text{s}$, $T_{\rm 2,ramsey} = 5.1 \pm 0.4~\mu\text{s}$, and $T_{\rm 2,echo} = 23.5 \pm 1.2~\mu\text{s}$, measured at the qubit's steepest point. (c) Qubit dephasing time $T_{\rm 2,\text{ramsey}}$ and resonance frequency test based on Ramsey interference, monitored over a 12-hour period. The frequency color fades as the qubit's frequency identification shifts into the noise background, representing $T_{\rm 2,ramsey}$. (d) Ramsy fringe frequency monitored over a 12-hour period. (e) Histogram of data from (d), showing frequency fluctuation within $\pm 40$~kHz (peak-to-peak $< 80$~kHz). (f) 12-hour $T_{\rm 2,ramsey}$ monitoring. (g) Histogram of data from (f), showing a Gaussian distribution of $T_{\rm 2,ramsey}$ between 3.5~$\sim$~5.5~$\mu$s.
    }
    \label{fig:5}
\end{figure*}

To ensure the precision of the 20-bit DAC, the reference voltage module is designed to deliver an ultra-low-noise, high stability output. The LTZ1000ACH is selected as the core reference device for its superior performance in precision voltage applications.~\scite{yang2022fpga, analog_devices_LTZ1000} It exhibits a noise density of $1.2\ \mu \mathrm{V}/\sqrt{\mathrm{Hz}}$ and a temperature coefficient of $0.05\ \mathrm{ppm}/^\circ \mathrm{C}$~\scite{analog_devices_LTZ1000}. These characteristics provide the necessary voltage stability and low noise required to maintain system accuracy. The LTZ1000ACH's low intrinsic noise also minimizes signal interference within the analog output chain.

The electronic performance of QPower is systematically characterized by benchmarking six key metrics: output ripple, root mean square (RMS) noise, low-frequency noise, high-frequency noise, long-term stability and crosstalk.

The output ripple, representing residual AC voltage superimposed on DC output, is quantified using a RIGOL DHO4804 oscilloscope in AC-coupled mode (20~MHz bandwidth limit). As shown in Fig.~\ref{fig:3}(a), the peak-to-peak ripple remains below 500~$\mu$V$_\mathrm{pp}$, with the instrument's intrinsic noise floor characterized as 100~$\mu$V$_\mathrm{pp}$ via 50~$\Omega$ termination. Output noise is further evaluated through RMS noise under identical configurations. Fig.~\ref{fig:3}(b) demonstrates RMS noise dependence on output voltage, peaking at 335~$\mu$V$_\mathrm{rms}$ for 7~V output. The oscilloscope's RMS noise floor, confirmed under a 50~$\Omega$ load, is approximately 80~$\mu$V$_\mathrm{rms}$.

To evaluate the noise performance comprehensively, low-frequency (10~Hz~$\sim$~100~kHz) and high-frequency (9~kHz~$\sim$~200~MHz) noise characteristics are measured using a Rohde \& Schwarz UPV audio analyzer and an FSL18 spectrum analyzer, respectively. As depicted in Fig.~\ref{fig:3}(c), the low-frequency noise spectral density at 10~kHz is approximately 20~nV/$\sqrt{\text{Hz}}$, with some peaks also attributable to the measurement system’s inherent noise (verified as 10~nV/$\sqrt{\text{Hz}}$ with a 50~$\Omega$ load). Fig.~\ref{fig:3}(d) shows the high-frequency spurious signals remain consistently below $-95$~dBm, with certain peaks originated from the analyzer itself.

Long-term stability is critical for frequency-tunable superconducting qubits, as DC voltage drift can lead to frequency shifts and coherence degradation. QPower employs instrumentation amplifier arrays combined with multi-stage positive and negative LDOs to minimize drift. Continuous stability measurements at three typical output voltages (0~V, 0.5~V, and 5~V) are performed over 12 hours using a KEITHLEY DMM6500 6.5-digit digital multimeter. As illustrated in Fig.~\ref{fig:3}(e), peak-to-peak voltage fluctuations at all measurement points remain below 5~$\mu$V\textsubscript{pp}. 

For multi-channel DC sources, channel-to-channel crosstalk is a critical concern, particularly in superconducting quantum circuits, where stringent electrical isolation is required to minimize interference between channels. To evaluate the channel-to-channel crosstalk within the same QPower module, the output of Channel 1 (CH1) is swept from $-7$~V to $+7$~V, while the output of Channel 2 (CH2) is monitored. A peak-to-peak deviation of less than 3~$\mu$V, corresponding to 0.3~ppm relative to the 14~V output range, is observed, as shown in Fig.~\ref{fig:3}(f).

\vspace{-6pt}
\section{Benchmark with superconducting qubits}
\vspace{-6pt}

To rigorously evaluate QPower's performance in practical quantum applications, we integrated QPower modules into the Microwave Measurement and Control System (M$^2$CS)~\scite{zhang2024m2cs} of a 66-qubit superconducting quantum processor~\scite{huang2025exact}. Specifically, three QPower modules are deployed: (1) DC biasing of a JPA at the mixing chamber stage, (2) simultaneous biasing of a cryogenic (4~K) high electron mobility transistor (HEMT) LNA and a room-temperature LNA, and (3) precise flux control for both a qubit and a tunable coupler, as illustrated in Fig.~\ref{fig:4}. The quantum processor and JPA are thermally anchored to the dilution refrigerator's mixing chamber, maintained at $<10\ \text{mK}$.

The qubit state measurement chain employs multi-stage filtering networks that suppress thermal noise interference across all cryogenic stages: RF-AWG generated probe signals undergo attenuation and filtering before entering the quantum chip, while the weak response signals experience sequential amplification through JPA (base temperature), cryogenic HEMT LNA, and room-temperature LNA prior to digitization by M$^2$CS's DAQ module. This architecture enables high-fidelity quantum state discrimination through noise-optimized signal conditioning.

QPower's stability directly impacts qubit coherence metrics -- we systematically evaluated energy relaxation time ($T_1$), Ramsey dephasing time ($T_{\rm 2,ramsey}$), and spin-echo coherence time ($T_{\rm 2,echo}$) at 3.8~GHz, an operating point that is most sensitive to external flux noise. As shown in Figs.~\ref{fig:5}(a-b), the measured $T_1$ reaches $87.6 \pm 3~\mu\text{s}$ s through $\pi$-pulse excited state decay measurements. Ramsey interferometry reveals $T_{\rm 2,ramsey} = 5.1 \pm 0.4~\mu\text{s}$, while spin-echo sequences extend coherence to $T_{\rm 2,echo} = 23.5 \pm 1.2~\mu\text{s}$ through low-frequency noise suppression~\scite{zhang2024m2cs}.  

Long-term stability was quantified through 12-hour continuous monitoring of the Ramsey interferometry, as shown in Fig.~\ref{fig:5}(c). Statistical analysis of 1000 repeated measurements reveals two key performance metrics: First, the Ramsey fringe exhibits $\pm 40$~kHz fluctuation, corresponding to peak-to-peak variations below 80~kHz in the qubit resonant frequency Figs.~\ref{fig:5}(d-e). Second, the Ramsey $T_{\rm 2,ramsey}$ maintains consistent performance within 3.5~$\sim$~5.5~\textmu s throughout the duration Figs.~\ref{fig:5}(f-g). These stability metrics demonstrate QPower's capability to maintain ppm level frequency control while preserving quantum coherence characteristics -- critical requirements for large-scale quantum computing systems.

\vspace{-6pt}

\section{Conclusion}
\vspace{-6pt}

We develop QPower, a low-noise, high-stability DC source tailored specifically for the demanding requirements of large-scale superconducting quantum processors. The system's hierarchical noise suppression architecture achieves benchmark performance metrics including: (1) voltage noise density $<20\ \text{nV}/\sqrt{\text{Hz}}$ at 10~kHz, and (2) long-term drift $<5\ \mu\text{V}$ over 12~hour operation -- parameters competitive with premium commercial instruments while consuming only 15~W per channel. This performance enables simultaneous support for multi-stage quantum measurement chains, including JPA biasing, cryogenic HEMT and room-temperature LNA powering, and qubit flux control in intermediate-scale quantum systems.

Experimental validation using a 66-qubit superconducting processor demonstrates QPower's capability to sustain quantum coherence characteristics matching state-of-the-art requirements: measured relaxation time $T_1 = 87.6 \pm 2~\mu\text{s}$ and Ramsey dephasing time $T_{\rm 2,ramsey} = 5.1 \pm 0.4~\mu\text{s}$ at the flux sensitive operating point. The system maintains qubit frequency stability with $\pm 40~\text{kHz}$ deviation during 12-hour continuous operation, directly correlating DC source noise performance with quantum coherence preservation -- a critical requirement for scalable quantum computing.

Furthermore, the modular design principle underlying QPower demonstrates extensibility to alternative quantum computing platforms, such as silicon-based quantum dot systems~\scite{he2019two,xue2022quantum,xue2021cmos}. This architectural flexibility positions QPower as a versatile solution for emerging quantum processor control architectures requiring distributed power systems, and high-precision measurement applications beyond quantum computing. 


\vspace{-6pt}

\begin{acknowledgments}
\vspace{-6pt}
{
 This work was supported by the Science, Technology and Innovation Commission of Shenzhen Municipality (KQTD20210811090049034), the Innovation Program for Quantum Science and Technology (2021ZD0301703).}
\end{acknowledgments}

\providecommand{\newblock}{}
\newcommand{\link}[2]{\href{#2}{#1}}
\bibliography{Q_POWER}

\begin{thebibliography}{42}%
\makeatletter
\providecommand \@ifxundefined [1]{%
 \@ifx{#1\undefined}
}%
\providecommand \@ifnum [1]{%
 \ifnum #1\expandafter \@firstoftwo
 \else \expandafter \@secondoftwo
 \fi
}%
\providecommand \@ifx [1]{%
 \ifx #1\expandafter \@firstoftwo
 \else \expandafter \@secondoftwo
 \fi
}%
\providecommand \natexlab [1]{#1}%
\providecommand \enquote  [1]{``#1''}%
\providecommand \bibnamefont  [1]{#1}%
\providecommand \bibfnamefont [1]{#1}%
\providecommand \citenamefont [1]{#1}%
\providecommand \href@noop [0]{\@secondoftwo}%
\providecommand \href [0]{\begingroup \@sanitize@url \@href}%
\providecommand \@href[1]{\@@startlink{#1}\@@href}%
\providecommand \@@href[1]{\endgroup#1\@@endlink}%
\providecommand \@sanitize@url [0]{\catcode `\\12\catcode `\$12\catcode
  `\&12\catcode `\#12\catcode `\^12\catcode `\_12\catcode `\%12\relax}%
\providecommand \@@startlink[1]{}%
\providecommand \@@endlink[0]{}%
\providecommand \url  [0]{\begingroup\@sanitize@url \@url }%
\providecommand \@url [1]{\endgroup\@href {#1}{\urlprefix }}%
\providecommand \urlprefix  [0]{URL }%
\providecommand \Eprint [0]{\href }%
\providecommand \doibase [0]{https://doi.org/}%
\providecommand \selectlanguage [0]{\@gobble}%
\providecommand \bibinfo  [0]{\@secondoftwo}%
\providecommand \bibfield  [0]{\@secondoftwo}%
\providecommand \translation [1]{[#1]}%
\providecommand \BibitemOpen [0]{}%
\providecommand \bibitemStop [0]{}%
\providecommand \bibitemNoStop [0]{.\EOS\space}%
\providecommand \EOS [0]{\spacefactor3000\relax}%
\providecommand \BibitemShut  [1]{\csname bibitem#1\endcsname}%
\let\auto@bib@innerbib\@empty
\bibitem [{\citenamefont {Castelvecchi}(2023)}]{IBM1000}%
  \BibitemOpen
  \bibfield  {author} {\bibinfo {author} {\bibfnamefont {D.}~\bibnamefont
  {Castelvecchi}},\ }\href@noop {} {\bibfield  {journal} {\bibinfo  {journal}
  {Nature}\ }\textbf {\bibinfo {volume} {624}},\ \bibinfo {pages} {238}
  (\bibinfo {year} {2023})}\BibitemShut {NoStop}%
\bibitem [{\citenamefont {Xu}\ \emph {et~al.}(2023{\natexlab{a}})\citenamefont
  {Xu}, \citenamefont {Sun}, \citenamefont {Wang}, \citenamefont {Xiang},
  \citenamefont {Bao}, \citenamefont {Zhu}, \citenamefont {Shen}, \citenamefont
  {Song}, \citenamefont {Zhang}, \citenamefont {Ren} \emph
  {et~al.}}]{xu2023digital}%
  \BibitemOpen
  \bibfield  {author} {\bibinfo {author} {\bibfnamefont {S.}~\bibnamefont
  {Xu}}, \bibinfo {author} {\bibfnamefont {Z.-Z.}\ \bibnamefont {Sun}},
  \bibinfo {author} {\bibfnamefont {K.}~\bibnamefont {Wang}}, \bibinfo {author}
  {\bibfnamefont {L.}~\bibnamefont {Xiang}}, \bibinfo {author} {\bibfnamefont
  {Z.}~\bibnamefont {Bao}}, \bibinfo {author} {\bibfnamefont {Z.}~\bibnamefont
  {Zhu}}, \bibinfo {author} {\bibfnamefont {F.}~\bibnamefont {Shen}}, \bibinfo
  {author} {\bibfnamefont {Z.}~\bibnamefont {Song}}, \bibinfo {author}
  {\bibfnamefont {P.}~\bibnamefont {Zhang}}, \bibinfo {author} {\bibfnamefont
  {W.}~\bibnamefont {Ren}}, \emph {et~al.},\ }\href@noop {} {\bibfield
  {journal} {\bibinfo  {journal} {Chinese Physics Letters}\ }\textbf {\bibinfo
  {volume} {40}},\ \bibinfo {pages} {060301} (\bibinfo {year}
  {2023}{\natexlab{a}})}\BibitemShut {NoStop}%
\bibitem [{\citenamefont {Acharya}\ \emph {et~al.}(2024)\citenamefont
  {Acharya}, \citenamefont {Abanin}, \citenamefont {Aghababaie-Beni},
  \citenamefont {Aleiner}, \citenamefont {Andersen}, \citenamefont {Ansmann},
  \citenamefont {Arute}, \citenamefont {Arya}, \citenamefont {Asfaw},
  \citenamefont {Astrakhantsev} \emph {et~al.}}]{acharya2024quantum}%
  \BibitemOpen
  \bibfield  {author} {\bibinfo {author} {\bibfnamefont {R.}~\bibnamefont
  {Acharya}}, \bibinfo {author} {\bibfnamefont {D.~A.}\ \bibnamefont {Abanin}},
  \bibinfo {author} {\bibfnamefont {L.}~\bibnamefont {Aghababaie-Beni}},
  \bibinfo {author} {\bibfnamefont {I.}~\bibnamefont {Aleiner}}, \bibinfo
  {author} {\bibfnamefont {T.~I.}\ \bibnamefont {Andersen}}, \bibinfo {author}
  {\bibfnamefont {M.}~\bibnamefont {Ansmann}}, \bibinfo {author} {\bibfnamefont
  {F.}~\bibnamefont {Arute}}, \bibinfo {author} {\bibfnamefont
  {K.}~\bibnamefont {Arya}}, \bibinfo {author} {\bibfnamefont {A.}~\bibnamefont
  {Asfaw}}, \bibinfo {author} {\bibfnamefont {N.}~\bibnamefont
  {Astrakhantsev}}, \emph {et~al.},\ }\href@noop {} {\bibfield  {journal}
  {\bibinfo  {journal} {Nature}\ } (\bibinfo {year} {2024})}\BibitemShut
  {NoStop}%
\bibitem [{\citenamefont {Gao}\ \emph {et~al.}(2025)\citenamefont {Gao},
  \citenamefont {Fan}, \citenamefont {Zha}, \citenamefont {Bei}, \citenamefont
  {Cai}, \citenamefont {Cai}, \citenamefont {Cao}, \citenamefont {Chen},
  \citenamefont {Chen}, \citenamefont {Chen} \emph
  {et~al.}}]{gao2025establishing}%
  \BibitemOpen
  \bibfield  {author} {\bibinfo {author} {\bibfnamefont {D.}~\bibnamefont
  {Gao}}, \bibinfo {author} {\bibfnamefont {D.}~\bibnamefont {Fan}}, \bibinfo
  {author} {\bibfnamefont {C.}~\bibnamefont {Zha}}, \bibinfo {author}
  {\bibfnamefont {J.}~\bibnamefont {Bei}}, \bibinfo {author} {\bibfnamefont
  {G.}~\bibnamefont {Cai}}, \bibinfo {author} {\bibfnamefont {J.}~\bibnamefont
  {Cai}}, \bibinfo {author} {\bibfnamefont {S.}~\bibnamefont {Cao}}, \bibinfo
  {author} {\bibfnamefont {F.}~\bibnamefont {Chen}}, \bibinfo {author}
  {\bibfnamefont {J.}~\bibnamefont {Chen}}, \bibinfo {author} {\bibfnamefont
  {K.}~\bibnamefont {Chen}}, \emph {et~al.},\ }\href@noop {} {\bibfield
  {journal} {\bibinfo  {journal} {Physical Review Letters}\ }\textbf {\bibinfo
  {volume} {134}},\ \bibinfo {pages} {090601} (\bibinfo {year}
  {2025})}\BibitemShut {NoStop}%
\bibitem [{\citenamefont {Ball}(2021)}]{ball2021first}%
  \BibitemOpen
  \bibfield  {author} {\bibinfo {author} {\bibfnamefont {P.}~\bibnamefont
  {Ball}},\ }\href@noop {} {\bibfield  {journal} {\bibinfo  {journal} {Nature}\
  }\textbf {\bibinfo {volume} {599}},\ \bibinfo {pages} {10} (\bibinfo {year}
  {2021})}\BibitemShut {NoStop}%
\bibitem [{\citenamefont {Choi}(2023)}]{choi2023ibm}%
  \BibitemOpen
  \bibfield  {author} {\bibinfo {author} {\bibfnamefont {C.~Q.}\ \bibnamefont
  {Choi}},\ }\href@noop {} {\bibfield  {journal} {\bibinfo  {journal} {IEEE
  Spectrum}\ }\textbf {\bibinfo {volume} {60}},\ \bibinfo {pages} {46}
  (\bibinfo {year} {2023})}\BibitemShut {NoStop}%
\bibitem [{\citenamefont {Liang}\ \emph {et~al.}(2018)\citenamefont {Liang},
  \citenamefont {Miao}, \citenamefont {Lin}, \citenamefont {Xu}, \citenamefont
  {Guo}, \citenamefont {Sun}, \citenamefont {Liao}, \citenamefont {Jin},\ and\
  \citenamefont {Peng}}]{liang2018ultra}%
  \BibitemOpen
  \bibfield  {author} {\bibinfo {author} {\bibfnamefont {F.}~\bibnamefont
  {Liang}}, \bibinfo {author} {\bibfnamefont {P.}~\bibnamefont {Miao}},
  \bibinfo {author} {\bibfnamefont {J.}~\bibnamefont {Lin}}, \bibinfo {author}
  {\bibfnamefont {Y.}~\bibnamefont {Xu}}, \bibinfo {author} {\bibfnamefont
  {C.}~\bibnamefont {Guo}}, \bibinfo {author} {\bibfnamefont {L.}~\bibnamefont
  {Sun}}, \bibinfo {author} {\bibfnamefont {S.}~\bibnamefont {Liao}}, \bibinfo
  {author} {\bibfnamefont {G.}~\bibnamefont {Jin}},\ and\ \bibinfo {author}
  {\bibfnamefont {C.}~\bibnamefont {Peng}},\ }\href@noop {} {\bibfield
  {journal} {\bibinfo  {journal} {arXiv preprint arXiv:1806.02645}\ } (\bibinfo
  {year} {2018})}\BibitemShut {NoStop}%
\bibitem [{\citenamefont {Zhang}\ \emph {et~al.}(2024)\citenamefont {Zhang},
  \citenamefont {Sun}, \citenamefont {Guo}, \citenamefont {Yuan}, \citenamefont
  {Zhang}, \citenamefont {Chu}, \citenamefont {Huang}, \citenamefont {Liang},
  \citenamefont {Qiu}, \citenamefont {Sun} \emph {et~al.}}]{zhang2024m2cs}%
  \BibitemOpen
  \bibfield  {author} {\bibinfo {author} {\bibfnamefont {J.}~\bibnamefont
  {Zhang}}, \bibinfo {author} {\bibfnamefont {X.}~\bibnamefont {Sun}}, \bibinfo
  {author} {\bibfnamefont {Z.}~\bibnamefont {Guo}}, \bibinfo {author}
  {\bibfnamefont {Y.}~\bibnamefont {Yuan}}, \bibinfo {author} {\bibfnamefont
  {Y.}~\bibnamefont {Zhang}}, \bibinfo {author} {\bibfnamefont
  {J.}~\bibnamefont {Chu}}, \bibinfo {author} {\bibfnamefont {W.}~\bibnamefont
  {Huang}}, \bibinfo {author} {\bibfnamefont {Y.}~\bibnamefont {Liang}},
  \bibinfo {author} {\bibfnamefont {J.}~\bibnamefont {Qiu}}, \bibinfo {author}
  {\bibfnamefont {D.}~\bibnamefont {Sun}}, \emph {et~al.},\ }\href@noop {}
  {\bibfield  {journal} {\bibinfo  {journal} {Chinese Physics B}\ }\textbf
  {\bibinfo {volume} {33}},\ \bibinfo {pages} {120309} (\bibinfo {year}
  {2024})}\BibitemShut {NoStop}%
\bibitem [{\citenamefont {Li}\ \emph {et~al.}(2019)\citenamefont {Li},
  \citenamefont {Yu}, \citenamefont {Tan}, \citenamefont {Zhao},\ and\
  \citenamefont {Yu}}]{li2019manipulation}%
  \BibitemOpen
  \bibfield  {author} {\bibinfo {author} {\bibfnamefont {Z.-Y.}\ \bibnamefont
  {Li}}, \bibinfo {author} {\bibfnamefont {H.-F.}\ \bibnamefont {Yu}}, \bibinfo
  {author} {\bibfnamefont {X.-S.}\ \bibnamefont {Tan}}, \bibinfo {author}
  {\bibfnamefont {S.-P.}\ \bibnamefont {Zhao}},\ and\ \bibinfo {author}
  {\bibfnamefont {Y.}~\bibnamefont {Yu}},\ }\href@noop {} {\bibfield  {journal}
  {\bibinfo  {journal} {Chinese Physics B}\ }\textbf {\bibinfo {volume} {28}},\
  \bibinfo {pages} {098505} (\bibinfo {year} {2019})}\BibitemShut {NoStop}%
\bibitem [{\citenamefont {Kalfus}\ \emph {et~al.}(2020)\citenamefont {Kalfus},
  \citenamefont {Lee}, \citenamefont {Ribeill}, \citenamefont {Fallek},
  \citenamefont {Wagner}, \citenamefont {Donovan}, \citenamefont {Rist{\`e}},\
  and\ \citenamefont {Ohki}}]{kalfus2020high}%
  \BibitemOpen
  \bibfield  {author} {\bibinfo {author} {\bibfnamefont {W.~D.}\ \bibnamefont
  {Kalfus}}, \bibinfo {author} {\bibfnamefont {D.~F.}\ \bibnamefont {Lee}},
  \bibinfo {author} {\bibfnamefont {G.~J.}\ \bibnamefont {Ribeill}}, \bibinfo
  {author} {\bibfnamefont {S.~D.}\ \bibnamefont {Fallek}}, \bibinfo {author}
  {\bibfnamefont {A.}~\bibnamefont {Wagner}}, \bibinfo {author} {\bibfnamefont
  {B.}~\bibnamefont {Donovan}}, \bibinfo {author} {\bibfnamefont
  {D.}~\bibnamefont {Rist{\`e}}},\ and\ \bibinfo {author} {\bibfnamefont
  {T.~A.}\ \bibnamefont {Ohki}},\ }\href@noop {} {\bibfield  {journal}
  {\bibinfo  {journal} {IEEE Transactions on Quantum Engineering}\ }\textbf
  {\bibinfo {volume} {1}},\ \bibinfo {pages} {1} (\bibinfo {year}
  {2020})}\BibitemShut {NoStop}%
\bibitem [{\citenamefont {Rietsche}\ \emph {et~al.}(2022)\citenamefont
  {Rietsche}, \citenamefont {Dremel}, \citenamefont {Bosch}, \citenamefont
  {Steinacker}, \citenamefont {Meckel},\ and\ \citenamefont
  {Leimeister}}]{rietsche2022quantum}%
  \BibitemOpen
  \bibfield  {author} {\bibinfo {author} {\bibfnamefont {R.}~\bibnamefont
  {Rietsche}}, \bibinfo {author} {\bibfnamefont {C.}~\bibnamefont {Dremel}},
  \bibinfo {author} {\bibfnamefont {S.}~\bibnamefont {Bosch}}, \bibinfo
  {author} {\bibfnamefont {L.}~\bibnamefont {Steinacker}}, \bibinfo {author}
  {\bibfnamefont {M.}~\bibnamefont {Meckel}},\ and\ \bibinfo {author}
  {\bibfnamefont {J.-M.}\ \bibnamefont {Leimeister}},\ }\href@noop {}
  {\bibfield  {journal} {\bibinfo  {journal} {Electronic Markets}\ }\textbf
  {\bibinfo {volume} {32}},\ \bibinfo {pages} {2525} (\bibinfo {year}
  {2022})}\BibitemShut {NoStop}%
\bibitem [{\citenamefont {Ryan}\ \emph {et~al.}(2017)\citenamefont {Ryan},
  \citenamefont {Johnson}, \citenamefont {Rist{\`e}}, \citenamefont {Donovan},\
  and\ \citenamefont {Ohki}}]{ryan2017hardware}%
  \BibitemOpen
  \bibfield  {author} {\bibinfo {author} {\bibfnamefont {C.~A.}\ \bibnamefont
  {Ryan}}, \bibinfo {author} {\bibfnamefont {B.~R.}\ \bibnamefont {Johnson}},
  \bibinfo {author} {\bibfnamefont {D.}~\bibnamefont {Rist{\`e}}}, \bibinfo
  {author} {\bibfnamefont {B.}~\bibnamefont {Donovan}},\ and\ \bibinfo {author}
  {\bibfnamefont {T.~A.}\ \bibnamefont {Ohki}},\ }\href@noop {} {\bibfield
  {journal} {\bibinfo  {journal} {Review of Scientific Instruments}\ }\textbf
  {\bibinfo {volume} {88}} (\bibinfo {year} {2017})}\BibitemShut {NoStop}%
\bibitem [{\citenamefont {C{\'o}rcoles}\ \emph {et~al.}(2019)\citenamefont
  {C{\'o}rcoles}, \citenamefont {Kandala}, \citenamefont {Javadi-Abhari},
  \citenamefont {McClure}, \citenamefont {Cross}, \citenamefont {Temme},
  \citenamefont {Nation}, \citenamefont {Steffen},\ and\ \citenamefont
  {Gambetta}}]{corcoles2019challenges}%
  \BibitemOpen
  \bibfield  {author} {\bibinfo {author} {\bibfnamefont {A.~D.}\ \bibnamefont
  {C{\'o}rcoles}}, \bibinfo {author} {\bibfnamefont {A.}~\bibnamefont
  {Kandala}}, \bibinfo {author} {\bibfnamefont {A.}~\bibnamefont
  {Javadi-Abhari}}, \bibinfo {author} {\bibfnamefont {D.~T.}\ \bibnamefont
  {McClure}}, \bibinfo {author} {\bibfnamefont {A.~W.}\ \bibnamefont {Cross}},
  \bibinfo {author} {\bibfnamefont {K.}~\bibnamefont {Temme}}, \bibinfo
  {author} {\bibfnamefont {P.~D.}\ \bibnamefont {Nation}}, \bibinfo {author}
  {\bibfnamefont {M.}~\bibnamefont {Steffen}},\ and\ \bibinfo {author}
  {\bibfnamefont {J.~M.}\ \bibnamefont {Gambetta}},\ }\href@noop {} {\bibfield
  {journal} {\bibinfo  {journal} {Proceedings of the IEEE}\ }\textbf {\bibinfo
  {volume} {108}},\ \bibinfo {pages} {1338} (\bibinfo {year}
  {2019})}\BibitemShut {NoStop}%
\bibitem [{\citenamefont {Wang}\ \emph {et~al.}(2021)\citenamefont {Wang},
  \citenamefont {Yu}, \citenamefont {Liu}, \citenamefont {Ma}, \citenamefont
  {Guo}, \citenamefont {Xiang}, \citenamefont {Song}, \citenamefont {Su},
  \citenamefont {Jin},\ and\ \citenamefont {Zheng}}]{wang_hardware_2021-1}%
  \BibitemOpen
  \bibfield  {author} {\bibinfo {author} {\bibfnamefont {Z.}~\bibnamefont
  {Wang}}, \bibinfo {author} {\bibfnamefont {H.}~\bibnamefont {Yu}}, \bibinfo
  {author} {\bibfnamefont {R.}~\bibnamefont {Liu}}, \bibinfo {author}
  {\bibfnamefont {X.}~\bibnamefont {Ma}}, \bibinfo {author} {\bibfnamefont
  {X.}~\bibnamefont {Guo}}, \bibinfo {author} {\bibfnamefont {Z.}~\bibnamefont
  {Xiang}}, \bibinfo {author} {\bibfnamefont {P.}~\bibnamefont {Song}},
  \bibinfo {author} {\bibfnamefont {L.}~\bibnamefont {Su}}, \bibinfo {author}
  {\bibfnamefont {Y.}~\bibnamefont {Jin}},\ and\ \bibinfo {author}
  {\bibfnamefont {D.}~\bibnamefont {Zheng}},\ }\href@noop {} {\bibfield
  {journal} {\bibinfo  {journal} {Chinese Physics B}\ }\textbf {\bibinfo
  {volume} {30}},\ \bibinfo {pages} {110305} (\bibinfo {year}
  {2021})}\BibitemShut {NoStop}%
\bibitem [{\citenamefont {Krantz}\ \emph {et~al.}(2019)\citenamefont {Krantz},
  \citenamefont {Kjaergaard}, \citenamefont {Yan}, \citenamefont {Orlando},
  \citenamefont {Gustavsson},\ and\ \citenamefont
  {Oliver}}]{krantz2019quantum}%
  \BibitemOpen
  \bibfield  {author} {\bibinfo {author} {\bibfnamefont {P.}~\bibnamefont
  {Krantz}}, \bibinfo {author} {\bibfnamefont {M.}~\bibnamefont {Kjaergaard}},
  \bibinfo {author} {\bibfnamefont {F.}~\bibnamefont {Yan}}, \bibinfo {author}
  {\bibfnamefont {T.~P.}\ \bibnamefont {Orlando}}, \bibinfo {author}
  {\bibfnamefont {S.}~\bibnamefont {Gustavsson}},\ and\ \bibinfo {author}
  {\bibfnamefont {W.~D.}\ \bibnamefont {Oliver}},\ }\href@noop {} {\bibfield
  {journal} {\bibinfo  {journal} {Applied physics reviews}\ }\textbf {\bibinfo
  {volume} {6}} (\bibinfo {year} {2019})}\BibitemShut {NoStop}%
\bibitem [{\citenamefont {Yang}\ \emph {et~al.}(2022)\citenamefont {Yang},
  \citenamefont {Shen}, \citenamefont {Zhu}, \citenamefont {Wang},
  \citenamefont {Zhang}, \citenamefont {Zhou}, \citenamefont {Jiang},
  \citenamefont {Deng},\ and\ \citenamefont {Liu}}]{yang2022fpga}%
  \BibitemOpen
  \bibfield  {author} {\bibinfo {author} {\bibfnamefont {Y.}~\bibnamefont
  {Yang}}, \bibinfo {author} {\bibfnamefont {Z.}~\bibnamefont {Shen}}, \bibinfo
  {author} {\bibfnamefont {X.}~\bibnamefont {Zhu}}, \bibinfo {author}
  {\bibfnamefont {Z.}~\bibnamefont {Wang}}, \bibinfo {author} {\bibfnamefont
  {G.}~\bibnamefont {Zhang}}, \bibinfo {author} {\bibfnamefont
  {J.}~\bibnamefont {Zhou}}, \bibinfo {author} {\bibfnamefont {X.}~\bibnamefont
  {Jiang}}, \bibinfo {author} {\bibfnamefont {C.}~\bibnamefont {Deng}},\ and\
  \bibinfo {author} {\bibfnamefont {S.}~\bibnamefont {Liu}},\ }\href@noop {}
  {\bibfield  {journal} {\bibinfo  {journal} {Review of Scientific
  Instruments}\ }\textbf {\bibinfo {volume} {93}} (\bibinfo {year}
  {2022})}\BibitemShut {NoStop}%
\bibitem [{\citenamefont {Ding}\ \emph {et~al.}(2024)\citenamefont {Ding},
  \citenamefont {Di~Federico}, \citenamefont {Hatridge}, \citenamefont {Houck},
  \citenamefont {Leger}, \citenamefont {Martinez}, \citenamefont {Miao},
  \citenamefont {I}, \citenamefont {Stefanazzi}, \citenamefont {Stoughton}
  \emph {et~al.}}]{ding2024experimental}%
  \BibitemOpen
  \bibfield  {author} {\bibinfo {author} {\bibfnamefont {C.}~\bibnamefont
  {Ding}}, \bibinfo {author} {\bibfnamefont {M.}~\bibnamefont {Di~Federico}},
  \bibinfo {author} {\bibfnamefont {M.}~\bibnamefont {Hatridge}}, \bibinfo
  {author} {\bibfnamefont {A.}~\bibnamefont {Houck}}, \bibinfo {author}
  {\bibfnamefont {S.}~\bibnamefont {Leger}}, \bibinfo {author} {\bibfnamefont
  {J.}~\bibnamefont {Martinez}}, \bibinfo {author} {\bibfnamefont
  {C.}~\bibnamefont {Miao}}, \bibinfo {author} {\bibfnamefont {D.~S.}\
  \bibnamefont {I}}, \bibinfo {author} {\bibfnamefont {L.}~\bibnamefont
  {Stefanazzi}}, \bibinfo {author} {\bibfnamefont {C.}~\bibnamefont
  {Stoughton}}, \emph {et~al.},\ }\href@noop {} {\bibfield  {journal} {\bibinfo
   {journal} {Physical Review Research}\ }\textbf {\bibinfo {volume} {6}},\
  \bibinfo {pages} {013305} (\bibinfo {year} {2024})}\BibitemShut {NoStop}%
\bibitem [{\citenamefont {Stefanazzi}\ \emph {et~al.}(2022)\citenamefont
  {Stefanazzi}, \citenamefont {Treptow}, \citenamefont {Wilcer}, \citenamefont
  {Stoughton}, \citenamefont {Bradford}, \citenamefont {Uemura}, \citenamefont
  {Zorzetti}, \citenamefont {Montella}, \citenamefont {Cancelo}, \citenamefont
  {Sussman}, \citenamefont {Houck}, \citenamefont {Saxena}, \citenamefont
  {Arnaldi}, \citenamefont {Agrawal}, \citenamefont {Zhang}, \citenamefont
  {Ding},\ and\ \citenamefont
  {Schuster}}]{stefanazziQICKQuantumInstrumentation2022}%
  \BibitemOpen
  \bibfield  {author} {\bibinfo {author} {\bibfnamefont {L.}~\bibnamefont
  {Stefanazzi}}, \bibinfo {author} {\bibfnamefont {K.}~\bibnamefont {Treptow}},
  \bibinfo {author} {\bibfnamefont {N.}~\bibnamefont {Wilcer}}, \bibinfo
  {author} {\bibfnamefont {C.}~\bibnamefont {Stoughton}}, \bibinfo {author}
  {\bibfnamefont {C.}~\bibnamefont {Bradford}}, \bibinfo {author}
  {\bibfnamefont {S.}~\bibnamefont {Uemura}}, \bibinfo {author} {\bibfnamefont
  {S.}~\bibnamefont {Zorzetti}}, \bibinfo {author} {\bibfnamefont
  {S.}~\bibnamefont {Montella}}, \bibinfo {author} {\bibfnamefont
  {G.}~\bibnamefont {Cancelo}}, \bibinfo {author} {\bibfnamefont
  {S.}~\bibnamefont {Sussman}}, \bibinfo {author} {\bibfnamefont
  {A.}~\bibnamefont {Houck}}, \bibinfo {author} {\bibfnamefont
  {S.}~\bibnamefont {Saxena}}, \bibinfo {author} {\bibfnamefont
  {H.}~\bibnamefont {Arnaldi}}, \bibinfo {author} {\bibfnamefont
  {A.}~\bibnamefont {Agrawal}}, \bibinfo {author} {\bibfnamefont
  {H.}~\bibnamefont {Zhang}}, \bibinfo {author} {\bibfnamefont
  {C.}~\bibnamefont {Ding}},\ and\ \bibinfo {author} {\bibfnamefont {D.~I.}\
  \bibnamefont {Schuster}},\ }\href@noop {} {\bibfield  {journal} {\bibinfo
  {journal} {Review of Scientific Instruments}\ }\textbf {\bibinfo {volume}
  {93}},\ \bibinfo {pages} {044709} (\bibinfo {year} {2022})}\BibitemShut
  {NoStop}%
\bibitem [{\citenamefont {Xu}\ \emph {et~al.}(2023{\natexlab{b}})\citenamefont
  {Xu}, \citenamefont {Huang}, \citenamefont {Fruitwala}, \citenamefont
  {Rajagopala}, \citenamefont {Naik}, \citenamefont {Nowrouzi}, \citenamefont
  {Santiago},\ and\ \citenamefont {Siddiqi}}]{xuQubiCExtensibleOpenSource2023}%
  \BibitemOpen
  \bibfield  {author} {\bibinfo {author} {\bibfnamefont {Y.}~\bibnamefont
  {Xu}}, \bibinfo {author} {\bibfnamefont {G.}~\bibnamefont {Huang}}, \bibinfo
  {author} {\bibfnamefont {N.}~\bibnamefont {Fruitwala}}, \bibinfo {author}
  {\bibfnamefont {A.}~\bibnamefont {Rajagopala}}, \bibinfo {author}
  {\bibfnamefont {R.~K.}\ \bibnamefont {Naik}}, \bibinfo {author}
  {\bibfnamefont {K.}~\bibnamefont {Nowrouzi}}, \bibinfo {author}
  {\bibfnamefont {D.~I.}\ \bibnamefont {Santiago}},\ and\ \bibinfo {author}
  {\bibfnamefont {I.}~\bibnamefont {Siddiqi}},\ }\href@noop {} {\bibinfo
  {title} {{{QubiC}} 2.0: {{An Extensible Open-Source Qubit Control System
  Capable}} of {{Mid-Circuit Measurement}} and {{Feed-Forward}}}} (\bibinfo
  {year} {2023}{\natexlab{b}})\BibitemShut {NoStop}%
\bibitem [{\citenamefont {Xu}\ \emph {et~al.}(2021)\citenamefont {Xu},
  \citenamefont {Huang}, \citenamefont {Balewski}, \citenamefont {Naik},
  \citenamefont {Morvan}, \citenamefont {Mitchell}, \citenamefont {Nowrouzi},
  \citenamefont {Santiago},\ and\ \citenamefont
  {Siddiqi}}]{xuQubiCOpenSourceFPGABased2021}%
  \BibitemOpen
  \bibfield  {author} {\bibinfo {author} {\bibfnamefont {Y.}~\bibnamefont
  {Xu}}, \bibinfo {author} {\bibfnamefont {G.}~\bibnamefont {Huang}}, \bibinfo
  {author} {\bibfnamefont {J.}~\bibnamefont {Balewski}}, \bibinfo {author}
  {\bibfnamefont {R.}~\bibnamefont {Naik}}, \bibinfo {author} {\bibfnamefont
  {A.}~\bibnamefont {Morvan}}, \bibinfo {author} {\bibfnamefont
  {B.}~\bibnamefont {Mitchell}}, \bibinfo {author} {\bibfnamefont
  {K.}~\bibnamefont {Nowrouzi}}, \bibinfo {author} {\bibfnamefont {D.~I.}\
  \bibnamefont {Santiago}},\ and\ \bibinfo {author} {\bibfnamefont
  {I.}~\bibnamefont {Siddiqi}},\ }\href@noop {} {\bibfield  {journal} {\bibinfo
   {journal} {IEEE Transactions on Quantum Engineering}\ }\textbf {\bibinfo
  {volume} {2}},\ \bibinfo {pages} {1} (\bibinfo {year} {2021})}\BibitemShut
  {NoStop}%
\bibitem [{\citenamefont {Guo}\ \emph {et~al.}(2019)\citenamefont {Guo},
  \citenamefont {Liang}, \citenamefont {Lin}, \citenamefont {Xu}, \citenamefont
  {Sun}, \citenamefont {Liu}, \citenamefont {Liao},\ and\ \citenamefont
  {Peng}}]{guoControlReadoutSoftware2019}%
  \BibitemOpen
  \bibfield  {author} {\bibinfo {author} {\bibfnamefont {C.}~\bibnamefont
  {Guo}}, \bibinfo {author} {\bibfnamefont {F.}~\bibnamefont {Liang}}, \bibinfo
  {author} {\bibfnamefont {J.}~\bibnamefont {Lin}}, \bibinfo {author}
  {\bibfnamefont {Y.}~\bibnamefont {Xu}}, \bibinfo {author} {\bibfnamefont
  {L.}~\bibnamefont {Sun}}, \bibinfo {author} {\bibfnamefont {W.}~\bibnamefont
  {Liu}}, \bibinfo {author} {\bibfnamefont {S.}~\bibnamefont {Liao}},\ and\
  \bibinfo {author} {\bibfnamefont {C.}~\bibnamefont {Peng}},\ }\href@noop {}
  {\bibfield  {journal} {\bibinfo  {journal} {IEEE Transactions on Nuclear
  Science}\ }\textbf {\bibinfo {volume} {66}},\ \bibinfo {pages} {1222}
  (\bibinfo {year} {2019})}\BibitemShut {NoStop}%
\bibitem [{\citenamefont {Lin}\ \emph {et~al.}(2019)\citenamefont {Lin},
  \citenamefont {Liang}, \citenamefont {Xu}, \citenamefont {Sun}, \citenamefont
  {Guo}, \citenamefont {Liao},\ and\ \citenamefont
  {Peng}}]{linScalableCustomizableArbitrary2019}%
  \BibitemOpen
  \bibfield  {author} {\bibinfo {author} {\bibfnamefont {J.}~\bibnamefont
  {Lin}}, \bibinfo {author} {\bibfnamefont {F.}~\bibnamefont {Liang}}, \bibinfo
  {author} {\bibfnamefont {Y.}~\bibnamefont {Xu}}, \bibinfo {author}
  {\bibfnamefont {L.-H.}\ \bibnamefont {Sun}}, \bibinfo {author} {\bibfnamefont
  {C.}~\bibnamefont {Guo}}, \bibinfo {author} {\bibfnamefont {S.-K.}\
  \bibnamefont {Liao}},\ and\ \bibinfo {author} {\bibfnamefont {C.-Z.}\
  \bibnamefont {Peng}},\ }\href@noop {} {\bibfield  {journal} {\bibinfo
  {journal} {AIP Advances}\ }\textbf {\bibinfo {volume} {9}},\ \bibinfo {pages}
  {115309} (\bibinfo {year} {2019})}\BibitemShut {NoStop}%
\bibitem [{\citenamefont {Sun}\ \emph {et~al.}(2020)\citenamefont {Sun},
  \citenamefont {Liang}, \citenamefont {Lin}, \citenamefont {Guo},
  \citenamefont {Xu}, \citenamefont {Liao},\ and\ \citenamefont
  {Peng}}]{sunScalableSelfAdaptiveSynchronous2020}%
  \BibitemOpen
  \bibfield  {author} {\bibinfo {author} {\bibfnamefont {L.}~\bibnamefont
  {Sun}}, \bibinfo {author} {\bibfnamefont {F.}~\bibnamefont {Liang}}, \bibinfo
  {author} {\bibfnamefont {J.}~\bibnamefont {Lin}}, \bibinfo {author}
  {\bibfnamefont {C.}~\bibnamefont {Guo}}, \bibinfo {author} {\bibfnamefont
  {Y.}~\bibnamefont {Xu}}, \bibinfo {author} {\bibfnamefont {S.}~\bibnamefont
  {Liao}},\ and\ \bibinfo {author} {\bibfnamefont {C.}~\bibnamefont {Peng}},\
  }\href@noop {} {\bibfield  {journal} {\bibinfo  {journal} {IEEE Transactions
  on Nuclear Science}\ }\textbf {\bibinfo {volume} {67}},\ \bibinfo {pages}
  {2148} (\bibinfo {year} {2020})}\BibitemShut {NoStop}%
\bibitem [{\citenamefont {Lisenfeld}\ \emph {et~al.}(2023)\citenamefont
  {Lisenfeld}, \citenamefont {Bilmes},\ and\ \citenamefont
  {Ustinov}}]{lisenfeld2023enhancing}%
  \BibitemOpen
  \bibfield  {author} {\bibinfo {author} {\bibfnamefont {J.}~\bibnamefont
  {Lisenfeld}}, \bibinfo {author} {\bibfnamefont {A.}~\bibnamefont {Bilmes}},\
  and\ \bibinfo {author} {\bibfnamefont {A.~V.}\ \bibnamefont {Ustinov}},\
  }\href@noop {} {\bibfield  {journal} {\bibinfo  {journal} {npj Quantum
  Information}\ }\textbf {\bibinfo {volume} {9}},\ \bibinfo {pages} {8}
  (\bibinfo {year} {2023})}\BibitemShut {NoStop}%
\bibitem [{\citenamefont {Grytsenko}\ \emph {et~al.}(2024)\citenamefont
  {Grytsenko}, \citenamefont {van Haagen}, \citenamefont {Rybalko},
  \citenamefont {Jennings}, \citenamefont {Mohan}, \citenamefont {Tian},\ and\
  \citenamefont {Kawakami}}]{grytsenko2024characterization}%
  \BibitemOpen
  \bibfield  {author} {\bibinfo {author} {\bibfnamefont {I.}~\bibnamefont
  {Grytsenko}}, \bibinfo {author} {\bibfnamefont {S.}~\bibnamefont {van
  Haagen}}, \bibinfo {author} {\bibfnamefont {O.}~\bibnamefont {Rybalko}},
  \bibinfo {author} {\bibfnamefont {A.}~\bibnamefont {Jennings}}, \bibinfo
  {author} {\bibfnamefont {R.}~\bibnamefont {Mohan}}, \bibinfo {author}
  {\bibfnamefont {Y.}~\bibnamefont {Tian}},\ and\ \bibinfo {author}
  {\bibfnamefont {E.}~\bibnamefont {Kawakami}},\ }\href@noop {} {\bibfield
  {journal} {\bibinfo  {journal} {arXiv preprint arXiv:2412.09811}\ } (\bibinfo
  {year} {2024})}\BibitemShut {NoStop}%
\bibitem [{\citenamefont {Terai}\ \emph {et~al.}(2003)\citenamefont {Terai},
  \citenamefont {Kameda}, \citenamefont {Yorozu}, \citenamefont {Fujimaki},\
  and\ \citenamefont {Wang}}]{terai2003effects}%
  \BibitemOpen
  \bibfield  {author} {\bibinfo {author} {\bibfnamefont {H.}~\bibnamefont
  {Terai}}, \bibinfo {author} {\bibfnamefont {Y.}~\bibnamefont {Kameda}},
  \bibinfo {author} {\bibfnamefont {S.}~\bibnamefont {Yorozu}}, \bibinfo
  {author} {\bibfnamefont {A.}~\bibnamefont {Fujimaki}},\ and\ \bibinfo
  {author} {\bibfnamefont {Z.}~\bibnamefont {Wang}},\ }\href@noop {} {\bibfield
   {journal} {\bibinfo  {journal} {IEEE transactions on applied
  superconductivity}\ }\textbf {\bibinfo {volume} {13}},\ \bibinfo {pages}
  {502} (\bibinfo {year} {2003})}\BibitemShut {NoStop}%
\bibitem [{\citenamefont {Macklin}\ \emph {et~al.}(2015)\citenamefont
  {Macklin}, \citenamefont {O’brien}, \citenamefont {Hover}, \citenamefont
  {Schwartz}, \citenamefont {Bolkhovsky}, \citenamefont {Zhang}, \citenamefont
  {Oliver},\ and\ \citenamefont {Siddiqi}}]{macklin2015near}%
  \BibitemOpen
  \bibfield  {author} {\bibinfo {author} {\bibfnamefont {C.}~\bibnamefont
  {Macklin}}, \bibinfo {author} {\bibfnamefont {K.}~\bibnamefont {O’brien}},
  \bibinfo {author} {\bibfnamefont {D.}~\bibnamefont {Hover}}, \bibinfo
  {author} {\bibfnamefont {M.}~\bibnamefont {Schwartz}}, \bibinfo {author}
  {\bibfnamefont {V.}~\bibnamefont {Bolkhovsky}}, \bibinfo {author}
  {\bibfnamefont {X.}~\bibnamefont {Zhang}}, \bibinfo {author} {\bibfnamefont
  {W.}~\bibnamefont {Oliver}},\ and\ \bibinfo {author} {\bibfnamefont
  {I.}~\bibnamefont {Siddiqi}},\ }\href@noop {} {\bibfield  {journal} {\bibinfo
   {journal} {Science}\ }\textbf {\bibinfo {volume} {350}},\ \bibinfo {pages}
  {307} (\bibinfo {year} {2015})}\BibitemShut {NoStop}%
\bibitem [{\citenamefont {Guo}\ \emph {et~al.}(2025)\citenamefont {Guo},
  \citenamefont {Sun}, \citenamefont {Huang}, \citenamefont {Sun},
  \citenamefont {Yuan}, \citenamefont {Zhang}, \citenamefont {Huang},
  \citenamefont {Liang}, \citenamefont {Qiu}, \citenamefont {Zhang} \emph
  {et~al.}}]{guo2025c}%
  \BibitemOpen
  \bibfield  {author} {\bibinfo {author} {\bibfnamefont {Z.}~\bibnamefont
  {Guo}}, \bibinfo {author} {\bibfnamefont {D.}~\bibnamefont {Sun}}, \bibinfo
  {author} {\bibfnamefont {P.}~\bibnamefont {Huang}}, \bibinfo {author}
  {\bibfnamefont {X.}~\bibnamefont {Sun}}, \bibinfo {author} {\bibfnamefont
  {Y.}~\bibnamefont {Yuan}}, \bibinfo {author} {\bibfnamefont {J.}~\bibnamefont
  {Zhang}}, \bibinfo {author} {\bibfnamefont {W.}~\bibnamefont {Huang}},
  \bibinfo {author} {\bibfnamefont {Y.}~\bibnamefont {Liang}}, \bibinfo
  {author} {\bibfnamefont {J.}~\bibnamefont {Qiu}}, \bibinfo {author}
  {\bibfnamefont {J.}~\bibnamefont {Zhang}}, \emph {et~al.},\ }\href@noop {}
  {\bibfield  {journal} {\bibinfo  {journal} {Chip}\ ,\ \bibinfo {pages}
  {100146}} (\bibinfo {year} {2025})}\BibitemShut {NoStop}%
\bibitem [{\citenamefont {Quinton}\ \emph {et~al.}(2025)\citenamefont
  {Quinton}, \citenamefont {Myhr}, \citenamefont {Barani}, \citenamefont
  {Crespo~del Granado},\ and\ \citenamefont {Zhang}}]{quinton2025quantum}%
  \BibitemOpen
  \bibfield  {author} {\bibinfo {author} {\bibfnamefont {F.~A.}\ \bibnamefont
  {Quinton}}, \bibinfo {author} {\bibfnamefont {P.~A.~S.}\ \bibnamefont
  {Myhr}}, \bibinfo {author} {\bibfnamefont {M.}~\bibnamefont {Barani}},
  \bibinfo {author} {\bibfnamefont {P.}~\bibnamefont {Crespo~del Granado}},\
  and\ \bibinfo {author} {\bibfnamefont {H.}~\bibnamefont {Zhang}},\
  }\href@noop {} {\bibfield  {journal} {\bibinfo  {journal} {Scientific
  Reports}\ }\textbf {\bibinfo {volume} {15}},\ \bibinfo {pages} {12733}
  (\bibinfo {year} {2025})}\BibitemShut {NoStop}%
\bibitem [{\citenamefont {Ladd}\ \emph {et~al.}(2010)\citenamefont {Ladd},
  \citenamefont {Jelezko}, \citenamefont {Laflamme}, \citenamefont {Nakamura},
  \citenamefont {Monroe},\ and\ \citenamefont {O’Brien}}]{ladd2010quantum}%
  \BibitemOpen
  \bibfield  {author} {\bibinfo {author} {\bibfnamefont {T.~D.}\ \bibnamefont
  {Ladd}}, \bibinfo {author} {\bibfnamefont {F.}~\bibnamefont {Jelezko}},
  \bibinfo {author} {\bibfnamefont {R.}~\bibnamefont {Laflamme}}, \bibinfo
  {author} {\bibfnamefont {Y.}~\bibnamefont {Nakamura}}, \bibinfo {author}
  {\bibfnamefont {C.}~\bibnamefont {Monroe}},\ and\ \bibinfo {author}
  {\bibfnamefont {J.~L.}\ \bibnamefont {O’Brien}},\ }\href@noop {} {\bibfield
   {journal} {\bibinfo  {journal} {nature}\ }\textbf {\bibinfo {volume}
  {464}},\ \bibinfo {pages} {45} (\bibinfo {year} {2010})}\BibitemShut
  {NoStop}%
\bibitem [{\citenamefont {{Analog Devices}}(2015)}]{analog_devices_LTZ1000}%
  \BibitemOpen
  \bibfield  {author} {\bibinfo {author} {\bibnamefont {{Analog Devices}}},\
  }\href@noop {} {\bibinfo {title} {{LTZ1000} {Datasheet} and {Product}
  {Info}}} (\bibinfo {year} {2015})\BibitemShut {NoStop}%
\bibitem [{\citenamefont {Huang}\ \emph {et~al.}(2025)\citenamefont {Huang},
  \citenamefont {Zhou}, \citenamefont {Zhang}, \citenamefont {Zhang},
  \citenamefont {Zhou}, \citenamefont {Guo}, \citenamefont {Yao}, \citenamefont
  {Huang}, \citenamefont {Li}, \citenamefont {Liang} \emph
  {et~al.}}]{huang2025exact}%
  \BibitemOpen
  \bibfield  {author} {\bibinfo {author} {\bibfnamefont {W.}~\bibnamefont
  {Huang}}, \bibinfo {author} {\bibfnamefont {X.-C.}\ \bibnamefont {Zhou}},
  \bibinfo {author} {\bibfnamefont {L.}~\bibnamefont {Zhang}}, \bibinfo
  {author} {\bibfnamefont {J.}~\bibnamefont {Zhang}}, \bibinfo {author}
  {\bibfnamefont {Y.}~\bibnamefont {Zhou}}, \bibinfo {author} {\bibfnamefont
  {Z.}~\bibnamefont {Guo}}, \bibinfo {author} {\bibfnamefont {B.-C.}\
  \bibnamefont {Yao}}, \bibinfo {author} {\bibfnamefont {P.}~\bibnamefont
  {Huang}}, \bibinfo {author} {\bibfnamefont {Q.}~\bibnamefont {Li}}, \bibinfo
  {author} {\bibfnamefont {Y.}~\bibnamefont {Liang}}, \emph {et~al.},\
  }\href@noop {} {\bibfield  {journal} {\bibinfo  {journal} {arXiv preprint
  arXiv:2502.19185}\ } (\bibinfo {year} {2025})}\BibitemShut {NoStop}%
\bibitem [{\citenamefont {Wang}\ and\ \citenamefont
  {Zhang}(2023)}]{wang2023design}%
  \BibitemOpen
  \bibfield  {author} {\bibinfo {author} {\bibfnamefont {W.}~\bibnamefont
  {Wang}}\ and\ \bibinfo {author} {\bibfnamefont {X.}~\bibnamefont {Zhang}},\
  }in\ \href@noop {} {\emph {\bibinfo {booktitle} {2023 5th International
  Conference on Electronic Engineering and Informatics (EEI)}}}\ (\bibinfo
  {organization} {IEEE},\ \bibinfo {year} {2023})\ pp.\ \bibinfo {pages}
  {584--588}\BibitemShut {NoStop}%
\bibitem [{\citenamefont {Egan}(2010)}]{egan201020}%
  \BibitemOpen
  \bibfield  {author} {\bibinfo {author} {\bibfnamefont {M.}~\bibnamefont
  {Egan}},\ }\href@noop {} {\bibfield  {journal} {\bibinfo  {journal} {Analog
  Dialogue}\ }\textbf {\bibinfo {volume} {44}},\ \bibinfo {pages} {1} (\bibinfo
  {year} {2010})}\BibitemShut {NoStop}%
\bibitem [{\citenamefont {de~Cos}\ \emph {et~al.}(2015)\citenamefont {de~Cos},
  \citenamefont {Su{\'a}rez}, \citenamefont {Garc{\i}} \emph
  {et~al.}}]{de2015hysteresis}%
  \BibitemOpen
  \bibfield  {author} {\bibinfo {author} {\bibfnamefont {J.}~\bibnamefont
  {de~Cos}}, \bibinfo {author} {\bibfnamefont {A.}~\bibnamefont {Su{\'a}rez}},
  \bibinfo {author} {\bibfnamefont {J.~A.}\ \bibnamefont {Garc{\i}}}, \emph
  {et~al.},\ }\href@noop {} {\bibfield  {journal} {\bibinfo  {journal} {IEEE
  Transactions on Microwave Theory and Techniques}\ }\textbf {\bibinfo {volume}
  {63}},\ \bibinfo {pages} {4284} (\bibinfo {year} {2015})}\BibitemShut
  {NoStop}%
\bibitem [{\citenamefont {Jenkins}(2013)}]{jenkins2013self}%
  \BibitemOpen
  \bibfield  {author} {\bibinfo {author} {\bibfnamefont {A.}~\bibnamefont
  {Jenkins}},\ }\href@noop {} {\bibfield  {journal} {\bibinfo  {journal}
  {Physics Reports}\ }\textbf {\bibinfo {volume} {525}},\ \bibinfo {pages}
  {167} (\bibinfo {year} {2013})}\BibitemShut {NoStop}%
\bibitem [{\citenamefont {Pandiev}(2023)}]{pandiev2023stability}%
  \BibitemOpen
  \bibfield  {author} {\bibinfo {author} {\bibfnamefont {I.~M.}\ \bibnamefont
  {Pandiev}},\ }\href@noop {} {\bibfield  {journal} {\bibinfo  {journal}
  {International Journal of Circuit Theory and Applications}\ }\textbf
  {\bibinfo {volume} {51}},\ \bibinfo {pages} {880} (\bibinfo {year}
  {2023})}\BibitemShut {NoStop}%
\bibitem [{\citenamefont {Safonov}\ and\ \citenamefont
  {Athans}(1977)}]{safonov1977gain}%
  \BibitemOpen
  \bibfield  {author} {\bibinfo {author} {\bibfnamefont {M.}~\bibnamefont
  {Safonov}}\ and\ \bibinfo {author} {\bibfnamefont {M.}~\bibnamefont
  {Athans}},\ }\href@noop {} {\bibfield  {journal} {\bibinfo  {journal} {IEEE
  Transactions on Automatic Control}\ }\textbf {\bibinfo {volume} {22}},\
  \bibinfo {pages} {173} (\bibinfo {year} {1977})}\BibitemShut {NoStop}%
\bibitem [{\citenamefont {Morroni}\ \emph {et~al.}(2009)\citenamefont
  {Morroni}, \citenamefont {Zane},\ and\ \citenamefont
  {Maksimovic}}]{morroni2009design}%
  \BibitemOpen
  \bibfield  {author} {\bibinfo {author} {\bibfnamefont {J.}~\bibnamefont
  {Morroni}}, \bibinfo {author} {\bibfnamefont {R.}~\bibnamefont {Zane}},\ and\
  \bibinfo {author} {\bibfnamefont {D.}~\bibnamefont {Maksimovic}},\
  }\href@noop {} {\bibfield  {journal} {\bibinfo  {journal} {IEEE Transactions
  on Power Electronics}\ }\textbf {\bibinfo {volume} {24}},\ \bibinfo {pages}
  {559} (\bibinfo {year} {2009})}\BibitemShut {NoStop}%
\bibitem [{\citenamefont {He}\ \emph {et~al.}(2019)\citenamefont {He},
  \citenamefont {Gorman}, \citenamefont {Keith}, \citenamefont {Kranz},
  \citenamefont {Keizer},\ and\ \citenamefont {Simmons}}]{he2019two}%
  \BibitemOpen
  \bibfield  {author} {\bibinfo {author} {\bibfnamefont {Y.}~\bibnamefont
  {He}}, \bibinfo {author} {\bibfnamefont {S.}~\bibnamefont {Gorman}}, \bibinfo
  {author} {\bibfnamefont {D.}~\bibnamefont {Keith}}, \bibinfo {author}
  {\bibfnamefont {L.}~\bibnamefont {Kranz}}, \bibinfo {author} {\bibfnamefont
  {J.}~\bibnamefont {Keizer}},\ and\ \bibinfo {author} {\bibfnamefont
  {M.}~\bibnamefont {Simmons}},\ }\href@noop {} {\bibfield  {journal} {\bibinfo
   {journal} {Nature}\ }\textbf {\bibinfo {volume} {571}},\ \bibinfo {pages}
  {371} (\bibinfo {year} {2019})}\BibitemShut {NoStop}%
\bibitem [{\citenamefont {Xue}\ \emph {et~al.}(2022)\citenamefont {Xue},
  \citenamefont {Russ}, \citenamefont {Samkharadze}, \citenamefont {Undseth},
  \citenamefont {Sammak}, \citenamefont {Scappucci},\ and\ \citenamefont
  {Vandersypen}}]{xue2022quantum}%
  \BibitemOpen
  \bibfield  {author} {\bibinfo {author} {\bibfnamefont {X.}~\bibnamefont
  {Xue}}, \bibinfo {author} {\bibfnamefont {M.}~\bibnamefont {Russ}}, \bibinfo
  {author} {\bibfnamefont {N.}~\bibnamefont {Samkharadze}}, \bibinfo {author}
  {\bibfnamefont {B.}~\bibnamefont {Undseth}}, \bibinfo {author} {\bibfnamefont
  {A.}~\bibnamefont {Sammak}}, \bibinfo {author} {\bibfnamefont
  {G.}~\bibnamefont {Scappucci}},\ and\ \bibinfo {author} {\bibfnamefont
  {L.~M.}\ \bibnamefont {Vandersypen}},\ }\href@noop {} {\bibfield  {journal}
  {\bibinfo  {journal} {Nature}\ }\textbf {\bibinfo {volume} {601}},\ \bibinfo
  {pages} {343} (\bibinfo {year} {2022})}\BibitemShut {NoStop}%
\bibitem [{\citenamefont {Xue}\ \emph {et~al.}(2021)\citenamefont {Xue},
  \citenamefont {Patra}, \citenamefont {van Dijk}, \citenamefont {Samkharadze},
  \citenamefont {Subramanian}, \citenamefont {Corna}, \citenamefont
  {Paquelet~Wuetz}, \citenamefont {Jeon}, \citenamefont {Sheikh}, \citenamefont
  {Juarez-Hernandez} \emph {et~al.}}]{xue2021cmos}%
  \BibitemOpen
  \bibfield  {author} {\bibinfo {author} {\bibfnamefont {X.}~\bibnamefont
  {Xue}}, \bibinfo {author} {\bibfnamefont {B.}~\bibnamefont {Patra}}, \bibinfo
  {author} {\bibfnamefont {J.~P.}\ \bibnamefont {van Dijk}}, \bibinfo {author}
  {\bibfnamefont {N.}~\bibnamefont {Samkharadze}}, \bibinfo {author}
  {\bibfnamefont {S.}~\bibnamefont {Subramanian}}, \bibinfo {author}
  {\bibfnamefont {A.}~\bibnamefont {Corna}}, \bibinfo {author} {\bibfnamefont
  {B.}~\bibnamefont {Paquelet~Wuetz}}, \bibinfo {author} {\bibfnamefont
  {C.}~\bibnamefont {Jeon}}, \bibinfo {author} {\bibfnamefont {F.}~\bibnamefont
  {Sheikh}}, \bibinfo {author} {\bibfnamefont {E.}~\bibnamefont
  {Juarez-Hernandez}}, \emph {et~al.},\ }\href@noop {} {\bibfield  {journal}
  {\bibinfo  {journal} {Nature}\ }\textbf {\bibinfo {volume} {593}},\ \bibinfo
  {pages} {205} (\bibinfo {year} {2021})}\BibitemShut {NoStop}%
\end{thebibliography}%
\bibliographystyle{apsrev4-2}

\end{document}